\documentclass[
superscriptaddress,
preprint,
showpacs,
amsmath,amssymb,
aps,
prb,
floatfix,
lengthcheck,
]{revtex4-1}

\usepackage{natbib}
\usepackage[dvipdfm]{graphicx,color}% Include figure files
\usepackage{dcolumn}% Align table columns on decimal point
\usepackage{bm}% bold math
\usepackage{ulem} % sout
\begin{document}

\title{Effects of frustration on magnetic excitations in a two-leg spin-ladder system}

\author{Takanori Sugimoto}
\email{sugimoto.takanori@jaea.go.jp}
\author{Michiyasu Mori}
\affiliation{Advanced Science Research Center, Japan Atomic Energy Agency, Tokai, Ibaraki 319-1195, Japan}
\affiliation{CREST, Japan Science and Technology Agency, Chiyoda-ku, Tokyo 102-0075, Japan}
\author{Takami Tohyama}
\affiliation{Yukawa Institute for Theoretical Physics, Kyoto University, Kyoto 606-8502, Japan}
\author{Sadamichi Maekawa}
\affiliation{Advanced Science Research Center, Japan Atomic Energy Agency, Tokai, Ibaraki 319-1195, Japan}
\affiliation{CREST, Japan Science and Technology Agency, Chiyoda-ku, Tokyo 102-0075, Japan}

\date{\today}

\begin{abstract}
We theoretically study the magnetic excitations in a frustrated two-leg spin-ladder system, in which antiferromagnetic exchange interactions act on the nearest-neighbor and next-nearest-neighbor bonds in the leg direction, and on the nearest-neighbor bonds in the rung direction.
A dynamical spin correlation function at zero temperature is calculated using the dynamical density-matrix renormalization-group method for possible magnetic phases, i.e., columnar-dimer and rung-singlet phases. 
The columnar-dimer phase is characterized by multi-spinon excitations with a spin gap, whereas the rung-singlet phase is dominated by the triplet excitation in the rung direction.
One major difference found between these two phases appears in the spectral weight of magnetic excitations, in particular, the bonding and anti-bonding modes in the rung direction. Therefore, we can distinguish one phase from the other by the difference in the spectral weight. 
Furthermore, we examine the effect of frustration on both modes in the rung-singlet phase with a perturbation theory from the strong coupling limit. 
The anti-bonding mode is shown to be stable against frustration, and a wave number with minimum excitation energy is shifted from being commensurate to being incommensurate. 
In contrast, the bonding mode is merged into the continuum excitation of multiple triplet excitations by increasing frustration.  
By comparing our results with inelastic neutron scattering experiments for BiCu$_{2}$PO$_{6}$, the magnitude of the magnetic exchange interactions and the ground state will be determined.
\end{abstract}

\pacs{75.10.Jm,75.10.Kt}
%75.10.Jm	Quantized spin models, including quantum spin frustration
%75.10.Kt	Quantum spin liquids, valence bond phases and related phenomena

\maketitle

%%%%%%%%%%%%%%%%%%%%%%%%
\section{Introduction}
%% spin-charge separation and spinon excitation %%
Spinons are elementary excitations characteristic of strongly correlated electron systems in low dimensions~\cite{maek04}, observable by angle-resolved photoemission spectroscopy~\cite{kim96} and neutron scattering~\cite{lake09}. 
There is a manifestation of electron fractionalization called spin-charge separation, which, instead of forming a single quasiparticle, the electronic excitation splits into spin (spinon) and charge (holon) degrees of freedom. 
In frustrated spin systems with exotic ground states such as spin-liquid and valence-bond-solid states, the spinon is associated with a topological excitation similar to the magnetic domain wall~\cite{bale10}. 
In contrast to the collective excitation in a magnetic long-ranged ordered state, i.e., magnon, the spinon carries a fractional quantum number, spin $1/2$, accompanied by a spin string. Because of this topological nature, two spinons make in some instances a bound state called a triplon~\cite{bale10,schm03}.

\begin{figure}[ht]
\includegraphics[scale=0.6]{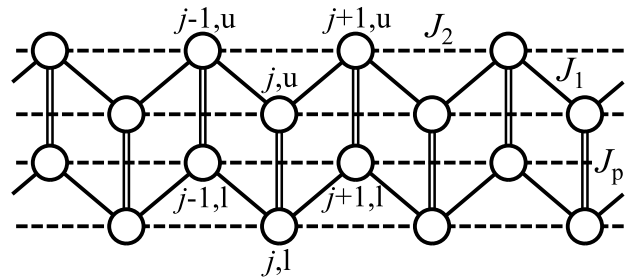}
\caption{\label{fig:tlzz} 
A frustrated two-leg spin-ladder system. Balls represent spin-$\frac{1}{2}$ sites. The nearest- and next-nearest-neighbor interactions in the leg direction, $J_1$ and $J_2$ are marked with solid and dashed lines, respectively. A double solid line denotes a nearest-neighbor interaction in the rung direction $J_p$.
}
\end{figure}

%% J1-J2 ladder %%
The frustrated two-leg spin-ladder system, i.e., the $J_1$-$J_2$-$J_p$ model (Fig.~\ref{fig:tlzz}), provides those exotic magnetic states, because this model bridges two different systems: the frustrated spin chain, i.e., $J_1$-$J_2$ model, and the non-frustrated spin ladder, i.e., $J_1$-$J_p$ model. 
Here, $J_1$ and $J_2$ are the nearest and the next-nearest antiferromagnetic exchange interactions, respectively, and $J_p$ is the antiferromagnetic interaction in the rung direction.
The ground-state phase diagram has been examined numerically~\cite{lavarelo11}, and has two phases: the columnar-dimer (CD) phase, which continuously merges to the dimer phase of the frustrated chain system ($J_1$-$J_2$ model) in the zero limit of $J_p$, and the rung-singlet (RS) phase, which continuously changes to the ground state of the non-frustrated ladder system ($J_1$-$J_p$ model) in the zero limit of $J_2$.
Note that the phase transition between the RS and the CD phases is of Ising type~\cite{lavarelo11}.
In contrast, the excited states of the $J_1$-$J_2$-$J_p$ model will be non-trivial as frustration leads to the confinement of spinons and triplons~\cite{bale10}. 
The two limiting cases of the $J_1$-$J_2$ and $J_1$-$J_p$ models are contained in the $J_1$-$J_2$-$J_p$ model. 
Therefore, this frustrated two-leg spin ladder is useful for understanding the effects of frustration on magnetic excitations. 
Several properties of the two limiting cases are summarized below.

%% J1-J2 chain %%
The $J_1$-$J_2$ model has two ground states: gapless spin-liquid and gapped dimer phases.
Between these phases, a Kosterlitz-Thouless transition occurs if the ratio of the exchange interactions, $J_2/J_1$, changes~\cite{affleck89,okamoto92,eggert96}.
In the spin-liquid phase, the spin correlation function exhibits a critical behavior and has no long-range order.
The excitations are well understood within the spinon picture, which is obtained by the Jordan-Wigner transformation of the spin operators.
In particular, the triplet excitation is dominated by two-spinon excitation~\cite{muller81,yokoyama97}.
In the dimer phase, the two neighboring spins constitute a singlet dimer, whose dimer correlation remains finite in the limit of infinite distance~\cite{majumdar69,white96}. 
The ground state is doubly degenerated owing to the two choices in forming a singlet dimer with either a right or a left neighboring site.

%% ladder %%
The intensive studies on the non-frustrated ladder system, i.e., the $J_1$-$J_p$ model, show that the one-dimensional spin-liquid phase is fragile with respect to the inter-chain interaction~\cite{barnes93}.
The RS phase appears in the ground state~\cite{barnes93}, where two spins connecting a rung bond constitute a singlet dimer and there is no degeneracy.
Thus, the RS phase is another gapped phase different from the dimer phase.
The magnetic excitation is dominated by the triplon, which is a bosonic quasiparticle defined as the triplet excitations from the RS dimers~\cite{gopalan94,sushkov98}.
It is also interpreted as a bound state of two spinons~\cite{lake09,bale10,schm03}.
The triplon always appears as an anti-bonding mode in the rung direction because of the parity changes between the ground state and the triplon excited state~\cite{haga02}.
An excitation with an even number of triplons appears as a bonding mode in the rung direction.
The bound states of two triplons are important for low-energy excitation in the bonding modes, because the intensity is much larger than the two-triplon continuum~\cite{sushkov98}.

%% purpose %%
In this paper, we investigate the magnetic excitations in the $J_1$-$J_2$-$J_p$ model.
Our main purpose is to clarify the effects of frustration by $J_2$ on the excitation spectra, which are observed by inelastic neutron scattering for BiCu$_2$PO$_6$ (Refs.~18-21). 
The compound should be compared with the unfrustrated two-leg spin-ladder compound, SrCu$_2$O$_3$ (Ref.~22). 
Not only the dispersion relation but also the spectral weight is crucial for analyzing experimental data.
Thus, we calculate the dynamical spin correlation function (DSCF) by the dynamical density-matrix renormalization-group (dynamical DMRG) method.
Furthermore, we propose how to determine the ground-state phase and how to estimate the interaction $J_1$, $J_2$, and $J_p$ in BiCu$_2$PO$_6$, by combining the DSCF and inelastic neutron scattering data.
Our supplementary purpose is to clarify the difference between the two phases, i.e., CD and RS phases, by the DSCF and to obtain the relationship between the exchange energies and the spectrum.

%% contents %%
The rest of the paper is organized as follows.
First, we present the model Hamiltonian of the frustrated two-leg spin-ladder system and method to calculate DSCF by dynamical DMRG in Sec.~\ref{sec:mam}.
We also establish notation for our perturbation theory treatment to explain excitation behaviors.
In Sec.~\ref{sec:res}, we present the numerical results and some interpretations of magnetic excitations.
We discuss the excitations from the perspective of frustration.
Ways to estimate the magnitude of the exchange interactions are also discussed.
Finally, we summarize the results and discuss consistency with previous studies in Sec.~\ref{sec:sad}.

%%%%%%%%%%%%%%%%%%%% 
\section{Model and method}
\label{sec:mam}
The model Hamiltonian of the frustrated two-leg spin-ladder system in Fig.~\ref{fig:tlzz} is defined as
\begin{equation}
\mathcal{H}=\sum_j J_p h_j^{(p)} + J_1 h_j^{(1)} + J_2 h_j^{(2)}
\label{eq:H}
\end{equation}
with
\begin{eqnarray}
h_j^{(p)}&=&\bm{S}_{j,\mathrm{u}}\cdot\bm{S}_{j,\mathrm{l}},\\
h_j^{(1)}&=&\bm{S}_{j,\mathrm{u}}\cdot\bm{S}_{j+1,\mathrm{u}}+\bm{S}_{j,\mathrm{l}}\cdot\bm{S}_{j+1,\mathrm{l}},\\
h_j^{(2)}&=&\bm{S}_{j,\mathrm{u}}\cdot\bm{S}_{j+2,\mathrm{u}}+\bm{S}_{j,\mathrm{l}}\cdot\bm{S}_{j+2,\mathrm{l}},
\end{eqnarray}
where $J_1(>0)$ and $J_2(>0)$ are the magnitudes of the antiferromagnetic nearest-neighbor and next-nearest-neighbor exchange interactions, respectively, and $J_p(>0)$ is that of the antiferromagnetic nearest-neighbor interaction in the rung direction.
$\bm{S}_{j,\mathrm{u(l)}}$ is the $S=1/2$ spin operator on the $j$ site in the upper (lower) chain.

To investigate magnetic excitations in the frustrated ladder system, we calculate the $z$-component of the DSCF at zero temperature given by
\begin{eqnarray}
\chi(\bm{q},\omega)&=&-\frac{1}{\pi} \Im \int_0^\infty dt e^{i\omega t} \langle 0| {S^z}^\dagger(\bm{q},t) S^z(\bm{q},0) |0\rangle \nonumber\\
&=&-\lim_{\gamma\to 0}\frac{1}{\pi} \Im \langle 0| S^z(-\bm{q})\frac{1}{\omega-\mathcal{H}+\epsilon_0+i\gamma} S^z(\bm{q}) |0\rangle,\nonumber \\
\label{eq:dscf}
\end{eqnarray}
where $|0\rangle$ is the ground state and $\epsilon_0$ is the ground-state energy.
The $x$ ($y$) component is equal to the $z$ component as the model Hamiltonian (\ref{eq:H}) displays no anisotropy. 
To obtain the excitation spectra, we set the infinitesimal value $\gamma$ to $0.1$, which is sufficient to examine the qualitative behaviors of the spin excitations in this system.
The momentum representation of the spin operator with the open boundary condition is given by
\begin{equation}
S^z(\bm{q})=\sqrt{\frac{2}{N+1}} \sum_j \sin(q_x j) S_{j,q_y}^z,
\end{equation}
with
\begin{eqnarray}
S_{j,q_y=0}^z&=&\frac{1}{2}(S_{j,\mathrm{u}}^z+S_{j,\mathrm{l}}^z), \\
S_{j,q_y=\pi}^z&=&\frac{1}{2}(S_{j,\mathrm{u}}^z-S_{j,\mathrm{l}}^z),
\end{eqnarray}
where the $x$ axis is aligned with the chain direction and the $y$ axis the rung direction.  
$N$ is the rung number of the system; specifically, the total number of spin sites is then $2N$.

We used the dynamical DMRG method~\cite{white93,jeckelmann02} to obtain the DSCF.
This method requires three target states: $|0\rangle$, $S^z(\bm{q})|0\rangle$, and $[\omega-\mathcal{H}+\epsilon_0+i\gamma]^{-1}S^z(\bm{q})|0\rangle$.
The last target state, the so-called correction vector, is obtained using a modified conjugate gradient method in the dynamical DMRG algorithm.

We checked the convergence of the ground-state energy for the DMRG truncation number $m$.
We confirmed that the ground-state energy per site converges within a numerical error of less than $10^{-3}$ for $m=150$ and used this value to obtain magnetic excitation spectra in a 32-rung frustrated ladder system.

A perturbation analysis based within the resonated valence bond (RVB) picture was also used to obtain the ground state and low-energy excited states.
We can rewrite the perturbed Hamiltonian $h_j^{(R)}$ ($R=1,2$) using projection operators onto RVB configurations.
The RVB picture helps us to handle complicated spin-configurations more easily (see the Appendix). 
We fix our notation for the singlet pair $[j,k]_p$ and the triplet with $S^z=0$, $\{j,k\}_p$ of two spins on the $j$ and $k$ sites ($j<k$) as follows:
\begin{eqnarray}
[j,k]_{p=\mathrm{i,u,l}}=\frac{1}{\sqrt{2}}(|\uparrow\rangle_{j,a_p}|\downarrow\rangle_{k,b_p}-|\downarrow\rangle_{j,a_p}|\uparrow\rangle_{k,b_p}), \\
\{j,k\}_{p=\mathrm{i,u,l}}=\frac{1}{\sqrt{2}}(|\uparrow\rangle_{j,a_p}|\downarrow\rangle_{k,b_p}+|\downarrow\rangle_{j,a_p}|\uparrow\rangle_{k,b_p}),
\end{eqnarray}
where the index $p$ denotes the position of singlet or triplet bond, i.e., $p=\mathrm{i}$ indicates an inter-chain bond, and $p=\mathrm{u (l)}$ signifies a bond in the upper (lower) chain. 
The local spin state on the $j$ site in the upper (lower) chain is given by $|\uparrow \rangle_{j,\mathrm{u (l)}}$ or $|\downarrow \rangle_{j,\mathrm{u (l)}}$. 
Thus, $a_{\mathrm{i}}=\mathrm{u}$ and $b_{\mathrm{i}}=\mathrm{l}$ for $p=\mathrm{i}$, whereas $a_{\mathrm{u}}=b_{\mathrm{u}}={\mathrm{u}}$ ($a_{\mathrm{l}}=b_{\mathrm{l}}={\mathrm{l}}$) for $p=\mathrm{u (l)}$.

Below, the index $p$ is omitted for the intra-chain, if the context is obvious.
We define two types of singlet configurations between the $j$ and $k$ rungs as
\begin{eqnarray}
\langle j,k\rangle_{\mathrm{r}}=[j,j]_{\mathrm{i}}[k,k]_{\mathrm{i}},\\
\langle j,k\rangle_{\mathrm{l}}=[j,k]_{\mathrm{u}}[j,k]_{\mathrm{l}}.
\end{eqnarray}
Using these expressions, the $S^z=0$ states in the total $S=0$ and $1$ subspaces with two triplets on the $j$ and $k$ rungs are obtained as
\begin{equation}
(j,k)_{0}=\frac{2}{\sqrt{3}}\left(\langle j,k\rangle_{\mathrm{l}}-\frac{1}{2}\langle j,k\rangle_{\mathrm{r}}\right)
\end{equation}
and
\begin{equation}
(j,k)_{1}=\frac{1}{\sqrt{2}}\bigg(\{j,k\}_{\mathrm{u}}[j,k]_{\mathrm{l}}+[j,k]_{\mathrm{u}}\{j,k\}_{\mathrm{l}}\bigg),
\end{equation}
respectively. 
To simplify the expressions developed from the perturbation theory, we use the Fourier transformation of the one- and two-triplet excited states given by
\begin{equation}
|1:q_x\rangle=\frac{1}{\sqrt{N}} \sum_j e^{iq_x j}\{j,j\}_{\mathrm{i}}\prod_{m\neq j}[m,m]_{\mathrm{i}}
\end{equation}
and
\begin{equation}
|2:S,(q_x,L)\rangle=\frac{1}{\sqrt{N}} \sum_j e^{iq_x j}(j,j+L)_{S}\prod_{m\neq j,j+L}[m,m]_{\mathrm{i}},
\end{equation}
respectively, where $S=0,1$ and $L=1,2,\cdots,(N-1)/2$.

%%%%%%%%%%%%%%%%%%%%%%
\section{Results}
\label{sec:res}
\begin{figure*}[ht]
\includegraphics[scale=1.2]{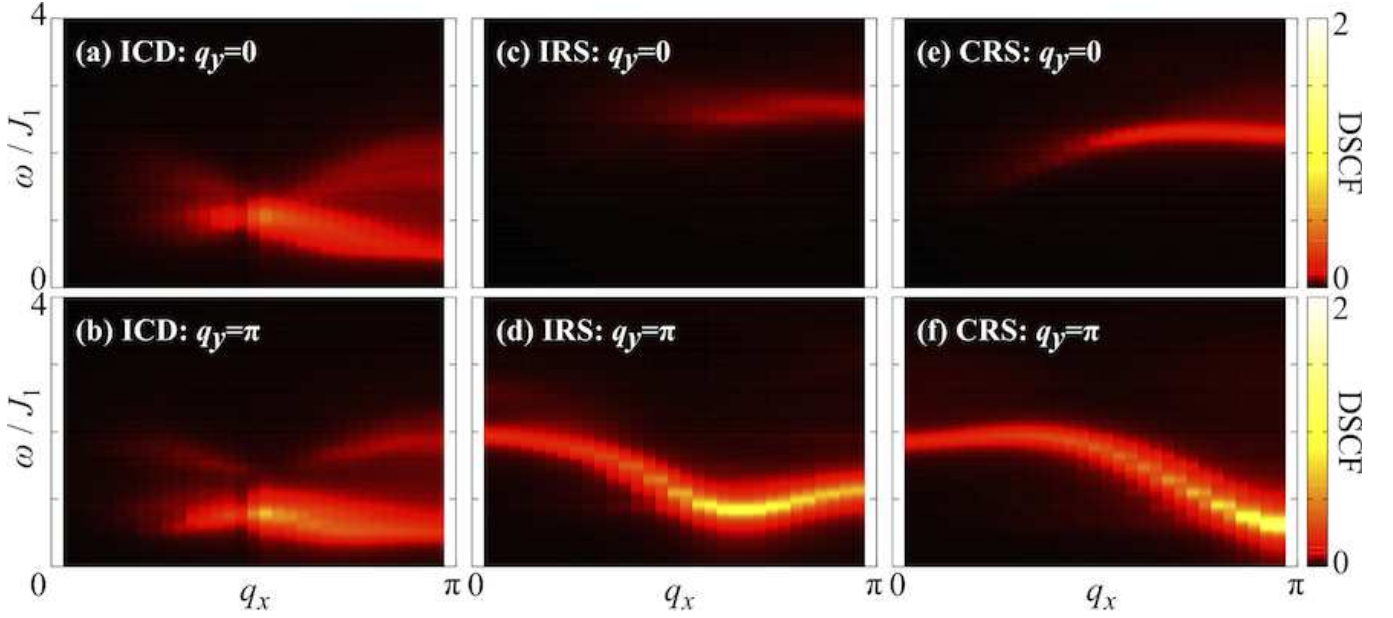}
\caption{\label{fig:DSCF} 
(Color online) The dynamical spin correlation functions (DSCF) of a 32-rung frustrated ladder for the three phases: (a), (b) the incommensurate CD phase ($J_p/J_1=0.2,\ J_2/J_1=0.6$), (c), (d) the incommensurate RS phase ($J_p/J_1=1.0,\ J_2/J_1=0.6$), and (e), (f) the commensurate RS phase ($J_p/J_1=1.0,\ J_2/J_1=0.1$). (a), (c), and (e) [(b), (d), and (f)] represent the excitations with $q_y=0$ ($q_y=\pi$) (Ref.~25). We set the DMRG truncation number $m=150$ and the broadening factor $\gamma=0.1$. 
}
\end{figure*}

In this section, we present numerical results of magnetic excitations in the frustrated two-leg spin-ladder system, and discuss the effects of frustration by $J_2$.
Figure \ref{fig:DSCF} shows the DSCF in a 32-rung frustrated ladder for three phases: the incommensurate CD phase ($J_p/J_1=0.2,\ J_2/J_1=0.6$), the incommensurate RS phase ($J_p/J_1=1.0,\ J_2/J_1=0.6$), and the commensurate RS phase ($J_p/J_1=1.0,\ J_2/J_1=0.1$).

First, we can find an obvious difference between the CD phase and the RS phase in the DSCF.
That is, the spectrum of the bonding mode in the CD phase [Fig.~\ref{fig:DSCF}(a)] exhibits almost the same shape with the same intensity as that of the anti-bonding mode [Fig.~\ref{fig:DSCF}(b)], whereas these are completely different in the RS phase [Figs.~\ref{fig:DSCF}(c)-\ref{fig:DSCF}(f)].
The difference between the CD phase and the RS phase is understood as the difference of the elementary excitations in both phases.
In the CD phase, the key elementary excitation is the spinon, which strongly reflects one dimensionality.
Thus, there are few differences between the bonding and anti-bonding modes in the CD phase.
By contrast, the major elementary excitation in the RS phase is the triplon, associated with excitations from the singlet to triplet state in rung bonds.
The parity of the triplet state is different from that of the singlet state.
Therefore, the excitation exchanging the rung parity, that is, the anti-bonding mode ($q_y=\pi$), is dominant in the RS phase.
With finite $J_1$, the ground state is mixed with singlet states in subspaces of even-number triplets, which has the same parity as the even-number triplet excitations.
Hence, the excitation of the bonding mode ($q_y=0$) appears with finite $J_1$.
We stress that the difference in the spectral weights plays a key role in identifying the ground-state phase using inelastic neutron scattering data for the frustrated two-leg spin-ladder compound, BiCu$_2$PO$_6$.

In the following, we clarify the origin of the spectra for each phase with three excitations: spinon, triplon, and bound state of two triplons.
The frustration effects are also clarified.

\subsection{Multi-spinon excitations}
The phase transition between the CD and RS phases in the frustrated ladder system belongs to the same universality class as the Ising transition~\cite{lavarelo11}.
The crucial point of the transition is a change of the ground-state degeneracy and the symmetry.
Although the ground state has no degeneracy in the RS phase, there are twofold-degenerate ground states in the CD phase.  
At zero temperature, the translation symmetry is spontaneously broken in the CD phase.
The CD phase continuously changes to the dimer phase in the $J_1$-$J_2$ model in the zero-coupling limit in the rung direction.
Therefore, the key excitation should be a spinon and is almost identical to the elementary excitation in the frustrated chain system.

In Figs.~\ref{fig:DSCF}(a) and \ref{fig:DSCF}(b), we can see the lower edge of the two-spinon continuum, which corresponds to the des Cloizeaux-Pearson mode in the non-frustrated limit $J_2\to 0$ in the chain system~\cite{yokoyama97,sugimoto12}.
The excitation has a minimum energy at an incommensurate wavenumber $q_x^{\ast}$ near the classical solution $q^{\ast}_{x(\mathrm{cl})}=\cos^{-1}(-J_1/4J_2)\cong q_x^{\ast}$.
The dispersion relation is almost symmetric with respect to $q_x=\pi/2$, although the intensity is weaker for $q_x<\pi/2$ than $q_x>\pi/2$.
This feature originates from the doubled unit cell because of the dimerization.
To explain this briefly, we consider the one-dimensional version of the DSCF for the frustrated chain system at the Majumdar-Ghosh point $J_2/J_1=0.5$ with the periodic boundary condition. 
The ground states,
\begin{eqnarray}
|\mathrm{MG}\rangle_{\mathrm{a}}=\prod_{j=0}^{N/2-1} [2j,2j+1], \\
|\mathrm{MG}\rangle_{\mathrm{b}}=\prod_{j=0}^{N/2-1} [2j+1,2j+2].
\end{eqnarray}
are doubly degenerate.
We assume that the system size $N$ is even.  
The ground-state energy of the Majumdar-Ghosh Hamiltonian is easily obtained as $\epsilon_{\mathrm{MG}}=-\frac{3}{8}NJ_1$.  
The one-dimensional version of the DSCF for a MG state $|\mathrm{MG}\rangle_{\mathrm{a}}$ is given by
\begin{equation}
\chi(q,\omega)^{(\mathrm{1D})}=-\frac{1-\cos(q)}{2\pi} \Im\langle q|\frac{1}{\omega-(\mathcal{H}_{\mathrm{MG}}-\epsilon_{\mathrm{MG}})+i0}|q\rangle,
\end{equation}
with
\begin{equation}
|q\rangle=\frac{1}{\sqrt{N}}\sum_{j=0}^{N/2-1} e^{2ijq} \{2j,2j+1\} \prod_{m\neq j} [2m,2m+1].
\end{equation}
Thus, we obtain the relation between $\chi(q,\omega)^{(\mathrm{1D})}$ and $\chi(\pi-q,\omega)^{(\mathrm{1D})}$ as
\begin{equation}
\chi(\pi-q,\omega)^{(\mathrm{1D})}=\frac{1+\cos(q)}{1-\cos(q)}\chi(q,\omega)^{(\mathrm{1D})},
\end{equation}
where $q\neq 0$. For $q=[\pi/2,\pi]$, we can easily confirm the relation regarding the spectral weight as, $\chi(\pi-q,\omega)^{(\mathrm{1D})}\leq \chi(q,\omega)^{(\mathrm{1D})}$ having the same poles.
This is due to the fact that the doubled periodicity of the dimer state, which occurs spontaneously, folds the Brillouin zone in half.
In other words, the original dispersion relation without dimerization and its inverted one with respect to $q=[\pi/2,\pi]$ are overlapped with each other in the Brillouin zone. 
Figures~\ref{fig:DSCF}(a) and \ref{fig:DSCF}(b) exemplify this instance.

\subsection{Triplon excitation}
We discuss the anti-bonding mode ($q_y=\pi$) in the RS phase [Figs.~\ref{fig:DSCF}(d) and \ref{fig:DSCF}(f)] in terms of the triplon.
The triplon is defined as the rung triplet excitation in this study. 
In perturbation theory under the strong coupling limit in the rung direction, the dispersion relation of the triplon to second order is obtained as
\begin{widetext} 
\begin{equation}
\epsilon_{\mathrm{T}}(q_x)=J_p\left[1+\alpha_1\cos(q_x)+\alpha_2\cos(2q_x)+\frac{3}{4}\left(\alpha_1^2+\alpha_2^2\right)+O(\alpha^3)\right], \label{eq:dr_trip}
\end{equation}
\end{widetext}
where $\alpha_1\equiv J_1/J_p$ and $\alpha_2\equiv J_2/J_p$~\cite{lavarelo11}.
The wave number at the minimum energy of the triplon excitation ${q_x^\ast}$ corresponds to the incommensurability.

In Fig.~\ref{fig:DSCF} (d), the wave number of the lowest excitation $q_x^\ast=(2/3\pm1/33)\pi$ is obtained for the $32$-rung ladder system.
The wave number is slightly shifted from the classical value ${q_x^\ast}_{(\mathrm{cl})}\cong 0.637\pi$.
It is known that the quantum correction to the wave number can be obtained by a mean-field analysis with rung-bond operators. 
Although demanding, the analysis requires proper renormalization to obtain a precise energy.  
Thus, the numerical calculation is indispensable in obtaining the precise wave number corresponding to an energy minimum in the dispersion relation of the triplon, especially for the large $\alpha_1$ and $\alpha_2$ regions.

In Fig.~\ref{fig:DSCF} (d), we find that the intensity is concentrated at the wave number $q \sim 2\pi/3$ associated with an energy minimum.
In Fig.~\ref{fig:DSCF} (f), both the intensity maximum and energy minimum shift from $q\sim2\pi/3$ to $q=\pi$ simultaneously.
Thus, we conclude that frustration affects only the wave number and the triplon picture remains stable.

\subsection{Bound state of two triplons}
In this section, we discuss the bonding mode ($q_y=0$) in the RS phase [Figs.~\ref{fig:DSCF}(c) and \ref{fig:DSCF}(e)].  
First, we show that without frustration $J_2=0$ the bonding mode disappears in the strong coupling limit $J_p/J_1\to\infty$ in the rung direction.
The ground state in this limit is given by the direct product of the rung singlets, $|0\rangle^{(0)}=\prod_j [j,j]_{\mathrm{i}}$.
In addition, the excited state at $q_y=0$ in Eq.~(\ref{eq:dscf}), $S^z(q_x,q_y=0)|0\rangle^{(0)}$, is explicitly rewritten as 
\begin{widetext}
\begin{equation}
S^z(q_x,0) |0\rangle^{(0)}=\Bigl( \frac{1}{\sqrt{N}} \sum_j e^{iq_x j}S_{j,q_y=0}^z \Bigr)
\Bigl( [j,j]_{\mathrm{i}} \prod_{m\neq j} [m,m]_{\mathrm{i}} \Bigr) = 0 \label{eq:sq0_r},
\end{equation}
\end{widetext}
because $S_{j,q_y=0}^z[j,j]_{\mathrm{i}} = 0$.
Therefore, the intensity of the bonding mode should vanish in the strong coupling limit in the rung direction.

\begin{figure}[ht]
\includegraphics[scale=0.6]{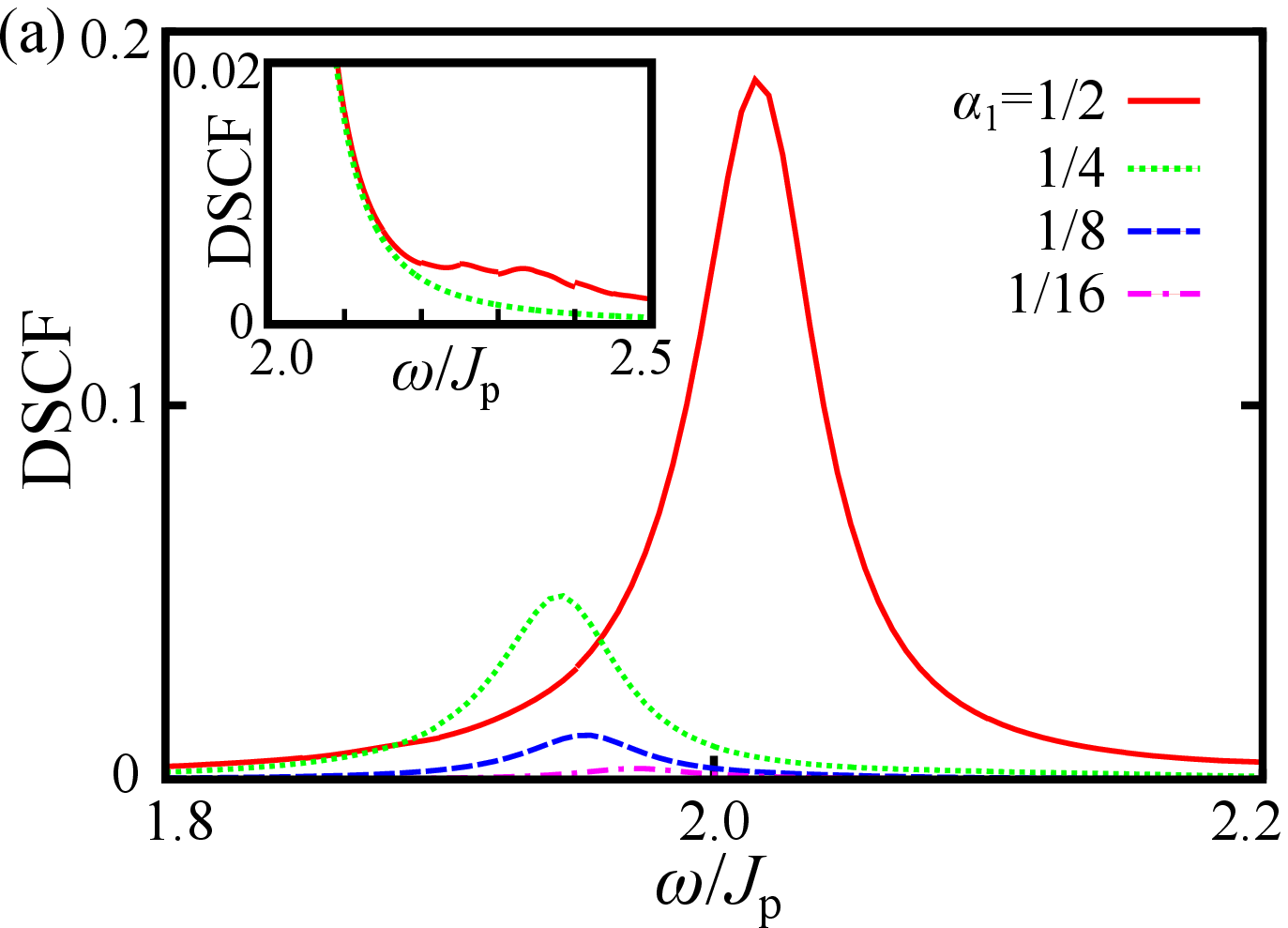}
\includegraphics[scale=0.6]{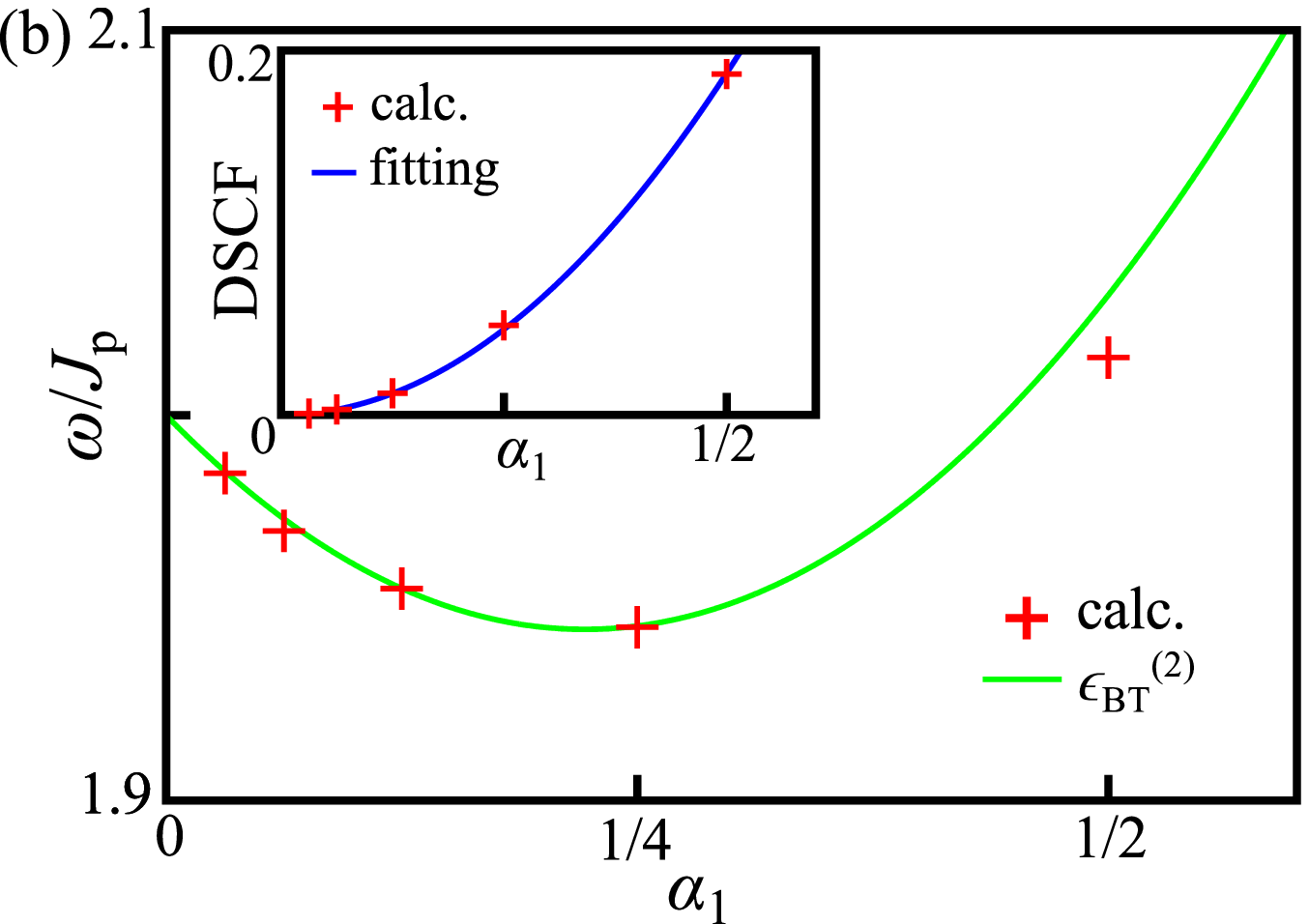}
\caption{\label{fig:qyeq0} 
(Color online) (a) $\alpha_1$ dependence of the DSCF in a 32-rung non-frustrated ladder, namely $J_2=0$, at $(q_x,q_y)=(32\pi/33,0)$ in the commensurate RS phase. 
The inset of (a) shows the DSCF for $\alpha_1=1/2$ (red solid line) and the Lorentzian fit (green dotted line).
(b) The energy of the peak vs $\alpha_1$ where the energy of the bound triplon with second-order perturbation theory, $\epsilon_{\mathrm{BT}}^{(2)}/J_p=2-\alpha_1/2+9\alpha_1^2/8$.
The inset of (b) shows the DSCF at the peak vs $\alpha_1$.
The blue line shows that the fitted function is proportional to $\alpha_1^2$ with the coefficient, $0.265\pm 0.002$.
We set the DMRG truncation number $m=300$ and the broadening factor $\gamma=0.05$ to distinguish the high-energy continuum from the main intensity of a single Lorentzian.
}
\end{figure}

Next, we consider a perturbation with of $\alpha_1=J_1/J_p\ll 1$ on $S=1$ excitation in the bonding mode.
The ground state within first-order perturbation is obtained as
\begin{equation}
|0\rangle^{(1)}=-\frac{\sqrt{3}}{4}\alpha_1\sqrt{N}|2:0,(0,1)\rangle. \label{eq:gs_1}
\end{equation}
The excited state from this state exists: $S^z(q_x,0)|0\rangle^{(1)}\neq 0$.
For example, the excited state at $q_x=\pi$ is obtained as
\begin{eqnarray}
S^z(\pi,0) |0\rangle^{(1)} = -\frac{\alpha_1}{\sqrt{2}} |2:1,(\pi,1)\rangle.
\end{eqnarray}
This is the origin of the bonding mode, the so-called {\it bound triplon}.
Since the bound triplon appears proportional to the first order of $\alpha_1$, the DSCF of the bound triplon starts from second order.
Figure \ref{fig:qyeq0} shows the $\alpha_1$ dependence of the bound triplon at $q_x=\pi$ obtained by the dynamical DMRG calculation for a 32-rung unfrustrated ladder with truncation number $m\geq300$.
In the inset of Fig.~\ref{fig:qyeq0} (b), we can see that the peak energy is proportional to $\alpha_1^2$.

We can obtain the excitation energy of the $S=1$ state of two triplons at $q_x=\pi$ by the perturbation theory.
In addition to the bound triplon whose energy is $\epsilon_{\mathrm{BT}}/J_p=2-\frac{1}{2}\alpha_1+\frac{9}{8}\alpha_1^2+O(\alpha_1^3)$ [Fig.~\ref{fig:qyeq0}(b)],  there is a continuum of two triplons starting from energy $\epsilon_{l}/J_p=2+\frac{9}{8}\alpha_1^2+O(\alpha_1^3)$.
Indeed, we can see the continuum above the bound triplon [the inset of Fig.~\ref{fig:qyeq0}(a)].
We note that the bound triplon is stabilized because of the dimerization energy in the leg, $-\frac{3}{4}J_1+\frac{1}{4}J_1=-\frac{1}{2}J_1$.

\begin{figure}[ht]
\includegraphics[scale=0.6]{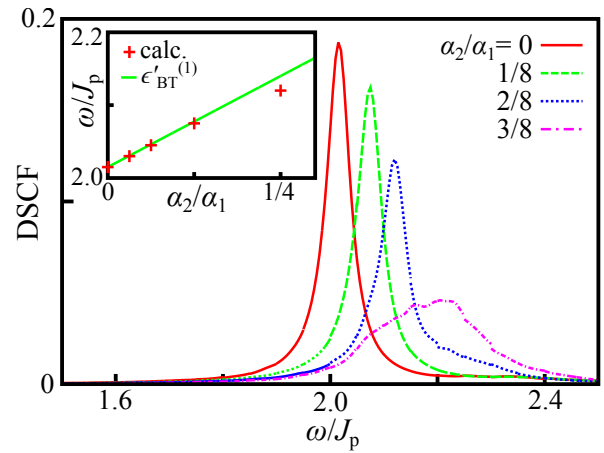}
\caption{\label{fig:j2dep} 
(Color online) $\alpha_2$ dependence of the DSCF in a 32-rung frustrated ladder at $(q_x,q_y)=(32\pi/33,0)$ and $\alpha_1=0.5$.
The inset shows the energy of the peak, corresponding to the energy of the bound triplon, as a function of $\alpha_2$.
Here, ${\epsilon_{\mathrm{BT}}^\prime}^{(1)}=\epsilon_{\mathrm{BT}}(\alpha_1)+\alpha_2 J_p$ (green line).
We set the DMRG truncation number $m=300$ and the broadening factor $\gamma=0.05$.
}
\end{figure}

The effect of frustration is also investigated using perturbation theory.
Figure \ref{fig:j2dep} shows the $\alpha_2$ dependence of DSCF at $q_x=\pi$ with $\alpha_1=0.5$. 
We can see that the energy of the bound triplon shifts to the continuum of two triplons as $\alpha_2$ increases.
At the same time, the weight of the bound triplon moves into the continuum.
To second order in $\alpha_1$ and first order in $\alpha_2$, i.e., $\alpha_2$/$\alpha_1$$\ll$1 is assumed, $\epsilon_{\mathrm{BT}}^\prime/J_p=2-\frac{1}{2}\alpha_1+\frac{9}{8}\alpha_1^2+\alpha_2+O(\alpha_1^3,\alpha_2^2,\alpha_1\alpha_2)$ at $q_x=\pi$ by the perturbation analysis.
In the inset of Fig.~\ref{fig:j2dep}, we confirm that the energy of the bound triplon agrees with the equation.
We expect the collective mode to be weakened and to be buried in the continuum with large $J_2/J_1$ in the strong coupling limit in the rung direction.  
Therefore, the bound triplon is unstable under large frustration, whereas the triplon is still stable with a slight shift in the wave number.

%%%%%%%%%%%%%%%%%%%%%%%%%%%%%
\section{Summary and discussion}
\label{sec:sad}
The frustrated two-leg spin-ladder system was investigated as a quantum spin system bridging two different systems: the frustrated chain system and the non-frustrated ladder system.
We examined the dynamical spin correlation function at zero temperature to understand the elementary excitations and the effects of frustration using the dynamical density-matrix renormalization-group method.
We analyzed the excitation spectra for three phases: the incommensurate columnar-dimer phase, the incommensurate rung-singlet phase and the commensurate rung-singlet phase.

First, we clarified that the $q_y$ dependence of the spectrum is different between the columnar-dimer and the rung-singlet phases.
This is because their elementary excitations are different between the two phases.
A key excitation in the columnar-dimer phase is the spinon, which is the same as in the frustrated chain system.  
In contrast, the main excitation in the rung-singlet phase is the triplon.  
The difference was explained using both the spinon picture for the columnar-dimer phase and the triplon picture for the rung-singlet phase.
With the help of perturbation theory under the strong coupling limit in the rung direction, we clarified the instability of the bound triplon with regard to frustration, although the triplon remains stable.

This is different from the effect of frustration on the triplon excitations in the ladder system~\cite{sushkov98}, where frustration is introduced with a diagonal antiferromagnetic interaction ($J_d$) in the nearest-neighbor-bond square.
This system corresponds to a frustrated chain with an alternating nearest-neighbor interaction~\cite{watanabe99,haga01,vekua06,note2}.
The RS phase appears without $J_d$, and the dimer phase appears at $J_d=J_1$ with large $J_1$ in the frustrated ladder system.
Therefore, from the point of view of the ground-state phase transition, the effect of in-chain frustration ($J_2$) is similar to that of diagonal frustration ($J_d$) for the ladder system.
However, frustration affects excitations differently, because the triplon excitation is fragile with respect to diagonal frustration, but not with in-chain frustration, as shown in the present study.
Thus, the effects of frustration on excitations are important for clarifying the influence of in-chain frustration in the ladder system.

Experimentally, this model will be useful for understanding the low-energy physics in the newly synthesized compound BiCu$_2$PO$_6$.
The magnetic structure of BiCu$_2$PO$_6$ is characterized by $\frac{1}{2}$ spins on Cu sites arranged in zigzag chains in the $b$ axis~\cite{koteswararao07,mentre09,tsirlin10}.
The zigzag chains are stacked up along the $c$ axis.
The exchange interactions $J_1$ and $J_2$ are defined as the nearest-neighbor and next-nearest-neighbor interactions in the zigzag chain.
In addition, there are comparable inter-chain interactions in BiCu$_2$PO$_6$.
As the distance between chains alternates, we define the rung exchange interactions $J_p$ and $J_p^\prime$ as the nearest-neighbor interactions on the shorter and longer bonds, respectively.
The compound was reported to be similar to a non-frustrated ladder system at first; that is, $J_2$ and $J_4$ are negligible as compared with $J_1$.
However, the numerical studies have shown that the susceptibility can be well understood as that for the frustrated two-leg spin-ladder system, where frustration is introduced as a next-nearest-neighbor antiferromagnetic interaction.
In particular, frustration introduced by $J_2$ should not be ignored.
Moreover, the band-structure calculation presented by Tsirlin {\it et al.}~\cite{tsirlin10} suggested that the $J_p^\prime$ interaction is comparable to $J_1$ and $J_2$, although $J_p$ is rather negligible.
With a comparison of our results and those from inelastic neutron scattering experiments~\cite{note1}, the magnitude of magnetic exchange interactions and the ground state of such a compound can be determined and will be reported in the near future.

{\it Note added.} Recently we noticed one paper on an inelastic neutron scattering experiment for BiCu$_2$PO$_6$~\cite{plumb13}. This data will be analyzed in light of our results.

%%%%%%%%%%%%%%%%%%%%%%%%%%%%%
\begin{acknowledgements}
We thank M. Fujita and R. Kajimoto for fruitful discussions.
Special thanks also go to T. Adachi, K. Ishii, K. Ikeuchi, K. Tsutsui, T. Sakai, and H. Onishi, who gave us invaluable comments.
This work was partly supported by Grants-in-Aid for Scientific Research from MEXT (Grants No.~24540387, No.~24360036, No.~23340093, and No.~22340097) and by the inter-university cooperative research program of IMR, Tohoku University.
We also acknowledge support from the Strategic Programs for Innovative Research (SPIRE), the Computational Materials Science Initiative (CMSI) and the Global COE Program ``The Next Generation of Physics, Spun from Universality and Emergence,'' Kyoto University.
Numerical computation in this work was performed on the supercomputers at JAEA, YITP in Kyoto University, and ISSP in The University of Tokyo.
\end{acknowledgements}

%%%%%%%%%%%%%%%%%%%%%%%%%%%%%
\appendix*
\section{PERTURBATION THEORY WITHIN RVB PICTURE}
We describe the perturbation-theory analysis in the rung-singlet phase for the two-leg spin-ladder system. 
Because of symmetries of the Hamiltonian, i.e., translation in the leg direction and reflection of two legs, the RVB picture is useful for analyzing low-energy states with perturbation theory.
Some simple relations help us to calculate the high-order states and the energies with perturbation theory.
In this paper, we concentrate on the ground state, one-triplet excitations, and two-triplet excitations.

First, we develop equations concerning to the rung-singlet configuration.
In the rung-singlet phase, the ground state is described as a linear combination of certain leg-singlet-pair states.
In practice, the perturbed Hamiltonian acting on the rung-singlet configuration generates a leg-singlet-pair state:
\begin{equation}
h_j^{(R)}\langle j,j+R\rangle_{\mathrm{r}}=\frac{1}{2}\langle j,j+R\rangle_{\mathrm{r}}-\langle j,j+R\rangle_{\mathrm{l}},
\end{equation}
where $R=1, 2$.
In addition, the perturbed Hamiltonian can generate a long-range leg-singlet pair and crossing leg-singlet pairs as follows,
\begin{widetext}
\begin{eqnarray}
h_j^{(R)}[j,j]_{\mathrm{i}}\langle j+R,k\rangle_{\mathrm{l}}&=&\frac{1}{2}[j+R,j+R]_{\mathrm{i}}\langle j,k\rangle_{\mathrm{l}}-\frac{1}{2}[k,k]_{\mathrm{i}} \langle j,j+R\rangle_{\mathrm{l}},\\
h_j^{(R)}\langle i,j\rangle_{\mathrm{l}}\langle j+R,k\rangle_{\mathrm{l}}&=&\frac{1}{2}\langle i,j+R\rangle_{\mathrm{l}}\langle j,k\rangle_{\mathrm{l}}-\frac{1}{2}\langle i,k\rangle_{\mathrm{l}} \langle j,j+R\rangle_{\mathrm{l}}.
\end{eqnarray}
\end{widetext}
In low-order perturbation, the corrections of the ground state and the excited states can be easily obtained with these equations. 

For the one-triplet excited state, the perturbed Hamiltonian can only move the triplet to another rung without generating leg-singlet pairs:
\begin{widetext}
\begin{eqnarray}
h_j^{(R)}\{j,j\}_{\mathrm{i}}[j+R,j+R]_{\mathrm{i}}&=&\frac{1}{2}[j,j]_{\mathrm{i}}\{j+R,j+R\}_{\mathrm{i}}, \label{eq:one-trip1}\\
h_j^{(R)}\{j,j\}_{\mathrm{i}}\langle j+R,k\rangle_{\mathrm{l}}&=&\frac{1}{2}\{j+R,j+R\}_{\mathrm{i}}\langle j,k\rangle_{\mathrm{l}}-\frac{1}{2}\{k,k\}_{\mathrm{i}} \langle j,j+R\rangle_{\mathrm{l}}.
\end{eqnarray}
\end{widetext}
These equations mean that frustration never makes a rung triplet unstable. 
In particular, Eq.~(\ref{eq:one-trip1}) determines the dispersion relation of the triplon in Eq.~(\ref{eq:dr_trip}).

In contrast, the perturbed Hamiltonian acting on a leg-triplet state generates a longer-range leg-triplet state,
\begin{widetext}
\begin{eqnarray}
h_j^{(R)}[j,j]_{\mathrm{i}}(j+R,k)_{1}&=&\frac{1}{2}[j+R,j+R]_{\mathrm{i}}(j,k)_{1},\\
h_j^{(R)}\langle i,j\rangle_{\mathrm{l}}(j+R,k)_{1}&=&\frac{1}{2}\langle i,j+R\rangle_{\mathrm{l}}(j,k)_{1}-\frac{1}{2}\langle j,j+R\rangle_{\mathrm{l}} (i,k)_{1}.
\end{eqnarray}
\end{widetext}
These imply that the bound state, i.e., a nearest-neighbor leg-triplet state, is no longer stable for any perturbation.
With only nearest-neighbor leg interaction $h_j^{(1)}$, off-diagonal parts of the perturbed Hamiltonian for the momentum distribution of leg-triplet state cancel at $q=\pi$.
This is the reason why the bound triplon is stable around $q=\pi$.
The next-nearest-neighbor leg interaction $h_j^{(2)}$ mixes the nearest-neighbor leg-triplet state with longer-range leg-triplet states.
Thus, the bound triplon is smeared out under large frustration.


\begin{thebibliography}{00}
%%% spin-charge separation and its observation by ARPES %%%
\bibitem{maek04} For a review, see S. Maekawa, T. Tohyama, S. E. Barnes, S. Ishihara, W. Koshibae, and G. Khaliullin, {\it Physics of Transition Metal Oxides}, Springer Series in Solid-State Sciences vol.~144 (Springer, Berlin, 2004). 
\bibitem{kim96} C. Kim, A. Y. Matsuura, Z.-X. Shen, N. Motoyama, H. Eisaki, S. Uchida, T. Tohyama, and S. Maekawa, Phys. Rev. Lett. {\bf 77}, 4054 (1996).
\bibitem{lake09} B. Lake, A. M. Tsvelik, S. Notbohm, D. A. Tennant, T. G. Perring, M. Reehuis, C. Sekar, G. Krabbes, and B. B\"uchner, Nature Physics {\bf 6}, 50 (2009).
\bibitem{bale10} For a review, see L. Balents, Nature (London) {\bf 464}, 199 (2010).
\bibitem{schm03} K. P. Schmidt and G. S. Uhrig, Phys. Rev. Lett. {\bf 90}, 227204 (2003).

%%% J1-J2 ladder %%%
\bibitem{lavarelo11} A. Lavar\'elo, G. Roux, and N. Laflorencie, Phys. Rev. B, {\bf 84}, 144407 (2011). %[BiCu2PO6, phaase diagram, dispersion (ED, triplon), suscep. (ED, magnon)]

%%% J1-J2 chain -- critical ratio %%%
\bibitem{affleck89} I. Affleck, D. Gepner, H. J. Schulz, and T. Ziman, J. Phys. A {\bf 22}, 511 (1989).
\bibitem{okamoto92} K. Okamoto and K. Nomura, Phys. Lett. A {\bf 169}, 433 (1992). 
\bibitem{eggert96} S. Eggert, Phys. Rev. B {\bf 54}, R9612 (1996).

%%% J1-J2 chain -- dynamics %%%
\bibitem{muller81} G. M\"uller, H. Thomas, H. Beck, and J. C. Bonner, Phys. Rev. B {\bf 24}, 1429 (1981).
\bibitem{yokoyama97} H. Yokoyama and Y. Saiga, J. Phys. Soc. Jpn. {\bf 66}, 3617 (1997).

%%% J1-J2 chain -- dimer %%%
\bibitem{majumdar69} C. K. Majumdar and D. K. Ghosh, J. Math. Phys. {\bf 10}, 1399 (1969).
\bibitem{white96} S. R. White and I. Affleck, Phys. Rev. B {\bf 54}, 9862 (1996).

%%% ladder %%%
\bibitem{barnes93} T. Barnes, E. Dagotto, J. Riera, and E. S. Swanson, Phys. Rev. B {\bf 47}, 3196 (1993). %[ladder, dispersion relation (spin wave, strong coupling, ED)]
\bibitem{gopalan94} S. Gopalan, T. M. Rice, and M. Sigrist, Phys. Rev. B {\bf 49}, 8901 (1994). %[ladder, DSCF (BOMF), triplon]
\bibitem{sushkov98} O. P. Sushkov and V. N. Kotov, Phys. Rev. Lett. {\bf 81}, 1941 (1998); V. N. Kotov, O. P. Sushkov, and R. Eder, Phys. Rev. B {\bf 59}, 6266 (1999). %[ladder, DSCF (BOMF), bound triplons]
\bibitem{haga02} N. Haga and S.-I. Suga, Phys. Rev. B {\bf 66}, 132415 (2002). %[ladder, DSCF (ED)]

%%% BCPO %%%
\bibitem{abraham94} F. Abraham, M. Ketatni, G. Mairesse, and B. Mernari, Eur. J. Solid State Inorg. Chem. {\bf 31}, 313 (1994). %[BiCu2PO6, crystal structure]
\bibitem{koteswararao07} B. Koteswararao, S. Salunke, A. V. Mahajan, I. Dasgupta, and J. Bobroff, Phys. Rev. B {\bf 76}, 052402 (2007). %[BiCu2PO6, suscep., spec. heat, DFT]
\bibitem{mentre09} O. Mentr\'e, E. Janod, P. Rabu, M. Hennion, F. Leclercq-Hugeux, J. Kang, C. Lee, M.-H. Whangbo, and S. Petit, Phys. Rev. B {\bf 80}, 180413(R) (2009). %[BiCu2PO6, suscep., INS, DFT]
\bibitem{tsirlin10} A. A. Tsirlin, I. Rousochatzakis, D. Kasinathan, O. Janson, R. Nath, F. Weickert, C. Geibel, A. M. L\"auchli, and H. Rosner, Phys. Rev. B {\bf 82}, 144426 (2010). %[BiCu2PO6, suscep., dispersion (triplon, magnon), J2 alternation, DFT]

%%% Experiments of INS for Ladder %%%
\bibitem{notbohm07} S. Notbohm, P. Ribeiro, B. Lake, D. A. Tennant, K. P. Schmidt, G. S. Uhrig, C. Hess, R. Klingeler, G. Behr, B. B{\"u}chner, M. Reehuis, R. I. Bewley, C. D. Frost, P. Manuel, and R. S. Eccleston, Phys. Rev. Lett. {\bf 98}, 027403 (2007). %[LaSrCuO, ladder, INS]

%%% DMRG %%%
\bibitem{white93} S. R. White, Phys. Rev. Lett. {\bf 69}, 2863 (1992); Phys. Rev. B {\bf 48}, 10345 (1993). %[DMRG]
\bibitem{jeckelmann02} E. Jeckelmann, Phys. Rev. B {\bf 66}, 045114 (2002). %[D-DMRG]

%%% note %%%
\bibitem{note1} We note that the Brillouin zone in the leg direction for BiCu$_2$PO$_6$ is half of our theoretical one because the unit cell in the compound includes two rungs.
Thus, the spectrum in Fig.~\ref{fig:DSCF} will be folded in half to be compared with experimental one.

%%% DSCF in J1-J2 chain %%%
\bibitem{sugimoto12} T. Sugimoto, S. Sota, and T. Tohyama, J. Phys. Soc. Jpn. {\bf 81}, 034706 (2012). %[DSCF, spin-Peierls system, quatum phonon]

%%% DSCF CuGeO3 (J1-J2) %%%
\bibitem{watanabe99} S. Watanabe and H. Yokoyama, J. Phys. Soc. Jpn. {\bf 68}, 2073 (1999). %[DSCF, family of J1-J2 chain]
\bibitem{haga01} N. Haga and S.-I. Suga, Phys. Rev. B {\bf 65}, 014414 (2001). %[DSCF, spin-Peierls sytem, field]

%%% ladder with diagonal frustration %%%
\bibitem{vekua06} T. Vekua and A. Honecker, Phys. Rev. B {\bf 73}, 214427 (2006).

%%% note %%%
\bibitem{note2} The interactions, $J_p$ and $J_d$, in the frustrated ladder system correspond to the alternating nearest-neighbor-interactions in the chain system, while $J_1$ corresponds to the next-nearest-neighbor one in the chain system.

%%% Experiments of INS for BCPO %%%
\bibitem{plumb13} K. W. Plumb, Z. Yamani, M. Matsuda, G. J. Shu, B. Koteswararao, F. C. Chou, and Y.-J. Kim, arXiv:1301.5324.

\end{thebibliography}
\end{document}